\documentstyle[11pt,newpasp,twoside,epsf]{article}
\def\edcomment#1{\iffalse\marginpar{\raggedright\sl#1\/}\else\relax\fi}
\reversemarginpar

\begin{document}
\title{Molecular line observations of proto-planetary nebulae}
\author{J. Alcolea, V. Bujarrabal, A. Castro-Carrizo, C. S\'anchez Contreras}
\affil{Observatorio Astron\'omico Nacional (OAN), Apartado 1143, E--28800 
Alcal\'a de Henares, Spain}
\author{R. Neri, J. Zweigle}
\affil{Institute de Radio Astronomie Milim\'etrique, 300 Rue de la Piscine, 
F--38406 St. Martin d'Heres, France}

\begin{abstract}

We present our recent results on mm-wave CO observations of proto-planetary 
nebulae.
These include high-resolution interferometric maps of various CO lines in
three well known bipolar PPNe: M\,1--92, M\,2--56 and OH\,231.8+4.2. The 
global properties of the high velocity molecular emission in post-AGB sources
have been also studied, by means of high-sensitivity single dish observations
of the $J$=1--0 and 2--1 lines of $^{12}$CO and $ ^{13}$CO. We discuss the
implications of these results to constrain the origin of the post-AGB
molecular high-velocity winds and the shaping of bipolar PPNe and PNe. In 
addition, we also present the results of an interferometric map of the 
molecular envelope around the luminous high-latitude star 89\,Her, a low mass 
post-AGB source which is also a close binary system.

\end{abstract}

\section{Introduction}

After attending this conference, I think that everyone will agree that the 
subject of the origin of asymmetrical planetary nebulae, in particular the
problem of the ``shaping'', is a matter of passionate debates, probably 
because our current observational knowledge of the problem still leads to too 
many possible solutions. From our point of view, this lack of knowledge is 
partially due to that planetary nebulae (PNe) themselves are not very good 
targets to study how this shaping process occurred. PNe are too old in two 
senses. On the one hand, regarding the problem of the shaping, we can consider
PNe as mainly consisting in freely expanding gas: the interactions presently 
active do not bear information on the large-scale shaping process (the origin 
of the axis-symmetry, etc.). On the other hand, when observing gas in PNe, it 
is very difficult to detect more than just a little fraction of the total mass 
of the progenitor AGB envelope: we do not probe the whole dynamics. 

One way to overcome these two difficulties 
is to study the objects in the phase between the AGB and PN stage, the 
proto-PNe (PPNe). Since they are younger, one expects them to show features of 
the more recent shaping processes, that could be even still active. They also 
have the advantage of being rich in molecular content, because the central 
star is not hot enough to drive a strong photo-ionization in its envelope. 
Due to this, mm-wave observations of the emission from CO rotational lines can 
efficiently trace the bulk of the mass in the envelope. Traditionally, this 
type of observations was of limited use because PPNe are rare (the PPN phase 
is very brief, it lasts less than 2\,000 yr), and therefore they are usually 
very far away and angularly small. Now, with instruments like the 
IRAM\footnote{IRAM is an European institute for research in millimeter 
astronomy, funded by the CNRS (France), the MPG (Germany), and the IGN 
(Spain)} 
Plateau de Bure (PdB) interferometer, we are able to image these sources with 
resolutions better than 1$''$, providing simultaneously spectral resolutions 
better than 1~km\,s$^{-1}$ and total mass sensitivities of the order of 
0.01~$M_{\sun}$. Here we summarize our results from the CO mapping
of several post-AGB envelopes using PdB: M\,1--92, M\,2--56, OH\,231.8+4.2, 
and 89\,Her (CO maps of CRL\,2688 are also presented by Lucas et al$.$ in this 
volume). For OH\,231.8+4.2 we have also mapped lines other than CO; these 
chemical studies are outlined in the contribution by S\'anchez Contreras et 
al$.$ In addition, we also present the preliminary results of a survey 
of CO lines in post-AGB sources, carried out with the IRAM 30~m radiotelescope.
We want to stress here the importance of this type of studies to understand
PN shaping. Note that every PN was once a PPN, and that whatever 
the properties of these PPNe were, models trying to reproduce the final stage 
of a certain PN should predict PPN stages similar to those we do observe.

\section{M\,1--92: Minkowski's footprint}

M\,1--92 is a bipolar PPN with a 20\,000~K central star, located 3~kpc away
and with a luminosity of 10$^4$ $L_{\sun}$, that was known to show high
velocity wings in molecular line emission. This is the object in which we 
started our mm-wave interferometric maps of PPNe, and therefore, the 
one we have studied in more detail. In addition to the high resolution 
maps of $^{12}$CO and $^{13}$CO we are discussing here, we have performed 
detailed imaging in the optical and NIR (using the WFPC2 and NICMOS2 on-board 
the HST), and continuum observations at 6 and 1.3~cm, and 2.6 and 1.3~mm 
(Alcolea et al$.$ 1999). 

The CO observations were conducted between 1993 and 1997, using the PdB 
interferometer, reaching a spatial resolution better than 1\farcs0 in the 
$^{13}$CO $J$=2--1 line (see Bujarrabal et al$.$ 1994, 1997, 1998a). The 
quality of the data and the good spatial resolution in comparison with the 
total size of the nebula, 10$''$, fully revealed the structure and kinematics 
of the molecular envelope, which consist of three main components (Fig.\,1a). 
The densest parts of the nebula lie in an equatorial disk (or torus) dividing 
the two reflecting lobes; this equatorial structure is expanding radially, at 
a constant AGB-like velocity of 8~km\,s$^{-1}$. At both sides of this 
component, the reflecting lobes are surrounded by two molecular shells, 
elongated along the symmetry axis of the nebula, and showing a constant 
velocity gradient (7.5~km\,s$^{-1}$ per $''$) in this direction. The walls
of these shells are very thin (0\farcs6): the two shells are practically empty.
The two lobes are ended by two high-velocity knots, moving at an expansion 
velocity of 70~km\,s$^{-1}$ along the axis. From the comparison of the 
different CO lines, we conclude that the kinetic temperature of the gas is 
very low, 10--30~K.

\begin{figure}
\plotfiddle{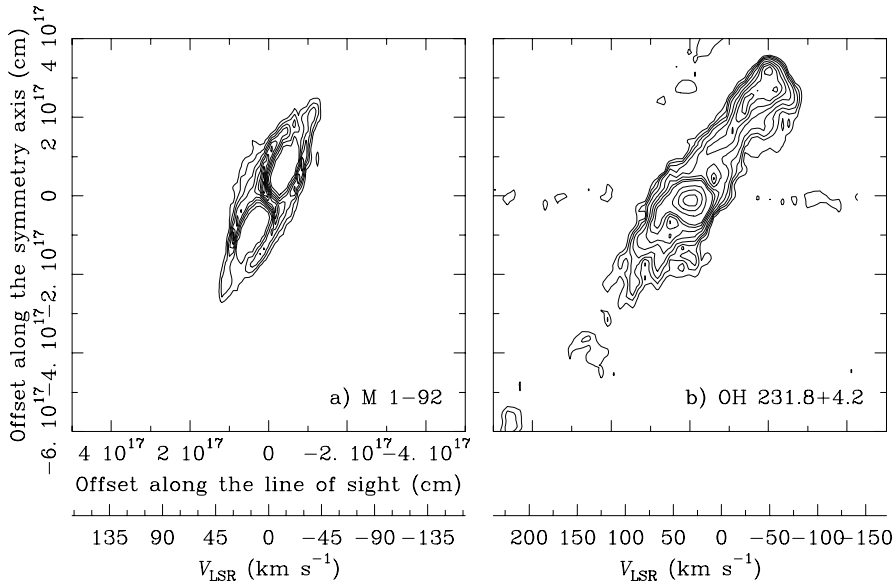}{7cm}{0}{80}{80}{-240}{-50}
\caption{Velocity ($V_{\rm LRS}$) vs. position 
(projected distance) diagrams for cuts along the symmetry axis for 
M\,1--92 (a) and OH\,231.8+4.2 (b), showing their 
characteristic Hubble-like velocity law. Because of this law, distances along 
the line of sight can be recovered by simply assuming that the gas is radially 
expanding at a velocity directly proportional to the distance to the central 
source, i.e. $\vec{V} \propto \vec{R}$. To visualize the structure of the 
nebula under this assumption use the scale in cm in both axes}
\end{figure}

The total mass of the neutral nebula detected by the CO observations is 
0.9~$M_{\sun}$, i.e., we are probably seeing the whole AGB envelope, strongly 
modified during the post-AGB evolution. We suggest that the present nebular 
configuration, which is only 900~yr 
old\footnote{This is the kinematic age, Age$_{\rm kin}$, derived from the 
``constant'' velocity gradient. This Hubble-like velocity field suggest that 
the interaction responsible for the present kinematics lasted much less 
than Age$_{\rm kin}$, i.e. $\la$~100--200 yr}, is due to some post-AGB mass 
ejection, highly bipolar and very fast, which impinged on two opposite sites 
of the AGB shell, accelerating and stretching it along the newly formed axis 
of symmetry, via a momentum (quasi-isothermal) driven shock. Assuming that 
during the AGB the nebula was spherically symmetric, and expanded at the 
velocity now only present in the thick equatorial disk, the axial momentum won 
by the envelope during the post-AGB phase is 3\,10$^{39}$~gr\,cm\,s$^{-1}$. 
This figure is nearly 10$^2$ times the amount of momentum available from 
photon pressure since the post-AGB interaction started. This result points 
out that not only the central star and the envelope change during the post-AGB 
evolution. The mass loss mechanism and its source of energy should also 
change, since photon pressure cannot account for the observed momentum. We 
have detected in SII and OI lines the present bipolar post-AGB flow in 
M\,1--92, as a chain of compact (0\farcs2) knots along the symmetry axis 
inside the CO empty shells (Bujarrabal et al$.$ 1998b); the momentum carried 
by this wind
is now very small, not affecting the kinematics of the envelope, but we cannot 
rule out that it were much larger in the past.

\section{M\,2--56}

M\,2--56 is also a PPN with a clear bipolar appearance in the optical, the
axis of symmetry being oriented in the East-West direction. Located 
$\sim$~3~kpc away, the central star shows a B0 (30\,000~K) spectral type and a 
luminosity of 10$^4$~$L_{\sun}$. In addition to PdB 4$''$ to 2$''$ resolution 
maps of the $^{12}$CO $J$=1--0 and 2--1 emission, we have also performed 
spectro-imaging of optical lines tracing shocked excited gas. 
Our CO observations are still not finished (there are 
some additional data to be taken) and therefore their analysis is very 
preliminary. However, the main results outlined here should not change 
(S\'anchez Contreras et al$.$ in prep.).

From the CO maps, M\,2--56 looks quite similar to M\,1--92, except that in the 
former, only the central parts of the nebula and the high velocity knots at 
both ends of the axis of symmetry are clearly detected. The two empty 
shells, characteristic of M\,1--92, appear weak and very 
fragmented in this case; they are mostly detected near the equatorial disk, 
which so resembles a sort of a flattened hourglass. Because of this 
lack of molecular emission between the equator and the poles of the nebula, 
and also because the symmetry axis lies almost in the plane of the sky, the 
constant velocity gradient is not so conspicuous here. However, when taking 
into account projection effects, we also obtain large values for the 
velocity gradient along the symmetry axis, 8.5~km\,s$^{-1}$ per $''$, the two 
high-velocity clumps escaping from the star at a velocity of 90~km\,s$^{-1}$. 
As for M\,1--92, the equatorial disk shows radial expansion with an 
AGB-like velocity (8--15~~km\,s$^{-1}$). We have measured a total envelope 
mass  of 0.1~$M_{\sun}$, the corresponding axial momentum being 
7\,10$^{38}$~gr\,cm\,s$^{-1}$. From the velocity gradient, we estimate that 
M\,2--56 is twice older than M\,1--92: Age$_{\rm kin}~\sim$ 1\,600--2\,000~yr. 
This result agrees with other age 
indicators that also point out that M\,2--56 should be more evolved than 
M\,1--92; the bluer spectral type and the weaker CO emission. In fact, the 
most likely explanation for the nearly absence of molecular emission in the 
lobe walls, is that, because of this aging, photo-dissociation by UV radiation 
has partially destroyed the molecular envelope. We conclude that M\,2--56 is 
an object very similar to 
M\,1--92, and therefore originated in a similar way, but somewhat more evolved.
In particular we suggest that the high-velocity knots detected in M\,2--56, 
which are detached from the rest of nebula, could be progenitors of FLIERs.

\begin{figure}
\plotone{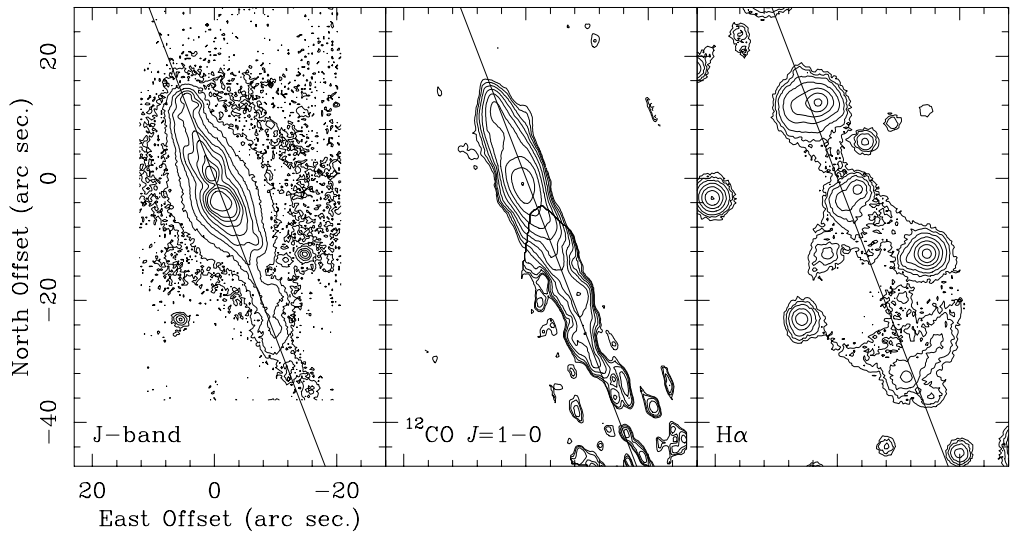}
\caption{The circumstellar envelope of OH\,231.8+4.2 as seen in the NIR 
(J-band, on the left) and in the total emission of the $^{12}$CO 
$J$=1--0 line (center), in comparison with the optical appearance 
(H$\alpha$+continuum, on the right). Note the different extent of the two 
lobes at all wavelengths, the similitude between the NIR and CO images, and 
the different shape of the southern lobe in the optical}
\end{figure}

\section{OH\,231.8+4.2}

OH\,231.8+4.2 is a very peculiar PPN. In the optical and NIR it shows a clear 
bipolar structure, but very asymmetric, the south lobe being more extended 
than the one in the north. The central star is also unique; located 
$\sim$~1.5~kpc away and with a luminosity of 10$^4$~$L_{\sun}$, it consists of 
a M9III Mira variable and maybe a hotter companion (it could be 
a symbiotic system). Not surprisingly, this nebula is also very peculiar in
molecular line emission, showing a very fast wind in CO and a particularly 
rich chemistry. In addition to PdB $^{12}$CO high resolution maps, we have 
also performed interferometric observations of other molecular species (see 
the contribution by S\'anchez Contreras et al$.$), and ground based optical 
long-slit observations and NIR imaging (Sanchez Contreras et al$.$ 1999; some 
beautiful NIR images from the HST NICMOS2 camera have been presented by 
Bieging et al$.$ in this conference). By the end of this year, we will also 
have HST WFPC2 images of the nebula in various atomic lines.

Our PdB interferometric maps of $^{12}$CO $J$=1--0 and 2--1 are not so detailed
as those of M\,1--92 (partly because the low declination of the source), but
are good enough as to show the structure of the nebula at a 2$''$ scale. The 
most outstanding feature of the molecular envelope is its shape; the CO 
nebula is extremely elongated, 1.6\,10$^{18}$~cm long (along the symmetry 
axis) but only 1.5\,10$^{17}$~cm across (taking into account an inclination 
of this axis
with respect to the line of sight of 50--55$\deg$). This distribution of 
matter agrees very well with the ground based NIR images, but strongly 
contrasts with the optical picture, especially in the south lobe (Fig.\,2). 
The kinematics of the molecular gas is dominated, once again, by the presence 
of a strong (8.9~km\,s$^{-1}$ per $''$, deprojected) and constant velocity 
gradient along the symmetry axis. We detect gas expanding up to 
375~km\,s$^{-1}$ in the south (210~km\,s$^{-1}$ in the north): to our  
knowledge, this is the highest expansion velocity ever measured in molecular 
line emission from evolved stars. The total mass of the nebula is 
0.5--1.0~$M_{\sun}$, and the axial momentum carried by the molecular flow, 
which is 750 yr old, is 3\,10$^{39}$~gr\,cm\,s$^{-1}$.

Although the most remarkable feature in the maps is the presence of the very
elongated fast emission in both lobes, the central core of the nebula is not
less important. Because of the somewhat lower resolution attained for  
OH\,231.8+4.2, the structure of the central parts of the nebula is not so 
clear. However, from the detailed inspection of the maps and its comparison 
with our first (also with 2$''$ resolution) observations of M\,1--92, we 
conclude that the two objects are very similar: there is a compact component 
at the very center of the nebula, and two hollow shells at both sides of it 
(Fig.\,1b; see also the discussion of the HCO$^+$ observations in the 
contribution by S\'anchez Contreras et al$.$). Therefore the main structural 
difference 
between the envelopes in M\,1--92, M\,2--56, and OH\,231+8+4.2 is that, in the 
latter, the two high-velocity knots have been replaced by two asymmetrical 
jets expanding a high velocity. In any case, we suggest that the origin of the
bipolarity of all three cases should be similar. The presence of the long 
high velocity CO jets in OH231.8+4.2 and not in the other two cases, could be 
due to differences in the details of the interaction responsible for the 
present shape (degree of collimation of the post-AGB jets, density contrast 
between the two colliding winds, etc.).

\section{Global properties of the high-velocity CO emission in PPNe}

The three objects just discussed provide many clues to investigate the
origin of bipolar PPNe and PNe, suggesting a similar formation mechanism in
all cases. However the sample is far from being statistically significant.
Because of the relatively small number of dishes per instrument and 
total collecting area, one of today's problems 
in high-quality imaging in mm-interferometry is that it is highly time 
consuming. Typically, high quality imaging, like that performed in any of the 
four targets here presented, takes about 1.5--2.5 full days of observing time. 
Given the large observing pressure 
in this type of instruments, to map a large (statistically significant)
sample of sources results prohibitive in these days. (This situation may 
change drastically if instruments like the planned ALMA become available.) To 
overcome these difficulties, we have carried out single dish observations, 
using the IRAM 30~m MRT, of a large number of PPNe, to study the global 
properties of the high-velocity molecular emission that characterizes this 
type of objects. The resolution of the telescope, 12$''$, does not allow 
to map the sources in almost all cases, but under certain assumptions, one can 
determine the mass, momentum, and kinetic energy of the high velocity outflow.
To better estimate the effects of line excitation and opacity, we have 
simultaneously observed the $J$=1--0 and 2--1 lines of $^{12}$CO and $^{13}$CO.
``Very preliminary'' results (Castro-Carrizo et al$.$ in prep.) of these 
observations are summarized in Table\,1.

In about 80\% of the studied PPNe we have found high-velocity CO wings, with
expansion velocities between 25 and 375~km\,s$^{-1}$. The derived total 
envelope mass can be as large as 0.5--1.0~$M_{\sun}$; in these cases it is 
clear that we have detected the whole circumstellar envelope. The 
measured axial momentum is always very large: the 
CO flows are ``hyper-luminous'' in the sense that the time required to obtain 
that momentum from photon pressure only (Age$_{L_\star}$), 
4\,000--100\,000 yr, is larger than the expected duration of the PPN phase. 
This problem becomes more severe if we consider the kinetic age 
(Age$_{\rm kin}$) of the flows in those objects in which it has been 
measured\footnote{Note that if our explanation for the Hubble-like velocity
laws found in these objects is correct, i.e., the momentum transfer occurred in
100--200 yr or less, the problem is even worst}. 
We note that the Age$_{L_\star}$ -- $P\rm_{axial}$/($L_\star$/$c$) -- does 
not depend on the distance assumed for the source, which is the main source 
of uncertainty in all these computations.

\begin{table}
\caption{Summary of the results from the IRAM 30~m observations for sources 
showing high velocity molecular emission. Figures are corrected for distance, 
opacity, and axial inclination. $V\rm_{exp}$ is the maximum expansion velocity
measured in CO; $M_{\rm CE}$ is the total mass of the molecular envelope; 
$P\rm_{axial}$ is the
linear momentum along the axial directions; Age$_{L_\star}$  is the time 
required for the photon pressure to account for the axial  momentum, i.e. 
$P\rm_{axial}$/($L_\star$/$c$); Age$_{\rm kin}$ is the kinetic age of the high 
velocity outflow when measured}
\smallskip
\begin{center}
\begin{tabular}{|c|cccccc|}\hline
Object & $V\rm_{exp}$ & $M_{\rm CE}$ & $P\rm_{axial}$ & $L_\star$ 
& Age$_{L_\star}$ & Age$_{\rm kin}$ \\
name & (km\,s$^{-1}$) & ($M_\odot$) & (gr\,cm/s)& ($L_\odot$)&(yr) &(yr)\\
\hline
OH\,231.8  & 375     & 1.0 & 3\,10$^{39}$   & ~\,\,10$^{4}$ & 70\,000 & ~\,\,750 \\
M\,1--92   & ~\,\,70 & 0.9 & 3\,10$^{39}$   & ~\,\,10$^{4}$ & 70\,000 & ~\,\,950 \\
M\,2--56   & ~\,\,90 & .07 & 7\,10$^{38}$ & ~\,\,10$^{4}$ & 30\,000 & 1\,800\\ \hline
Frosty Leo & 235     & .09 & 9\,10$^{38}$   & 3\,10$^{3}$   & 80\,000 & \\
He\,3--1475& ~\,\,70 & .07 & 3\,10$^{38}$   & ~\,\,10$^{3}$ & 50\,000 & \\
IRC\,+10420& ~\,\,50 & 2.0 & ~\,\,10$^{40}$ & 7\,10$^{5}$   & ~\,5\,000 & \\
CRL~2688   & 190     & 0.7 & 5\,10$^{39}$   & 4\,10$^{4}$   & 30\,000 & \\
CRL~618    & 240     & 0.6 & 2\,10$^{39}$   & 3\,10$^{4}$   & 10\,000 & \\
IRAS\,1743 & ~\,\,28 & 0.6 & 2\,10$^{39}$   & 6\,10$^{4}$   & ~\,6\,000 & \\
IRAS\,1911 & ~\,\,44 & .04 & 2\,10$^{39}$   & 3\,10$^{4}$   & 20\,000 & \\
IRAS\,1950 & ~\,\,55 & .03 & 8\,10$^{37}$   & 2\,10$^{3}$   & 10\,000 & \\
IRAS\,2227 & ~\,\,25 & .07 & ~\,\,10$^{38}$ & 8\,10$^{3}$   & ~\,4\,000 & \\ \hline
\end{tabular}
\end{center}
\end{table}

These results, in addition to those obtained from the high resolution maps of
PPNe (bipolarity, relative high degree of collimation, absence of large scale
rotation in the envelopes), impose very strong constrains for the possible 
mechanisms capable of powering these molecular ejections, that we believe are 
also the responsible for the shaping of bipolar and point symmetric PNe. In 
fact, the large momentum and energy released in just a few hundred years rule 
out not only photon pressure, but probably all other mechanisms not powered by 
the accretion of material or by explosive nuclear reactions. 
On the other hand, 
the similitude between the post-AGB high-velocity molecular flows and those in 
young stellar objects, suggests that maybe also here the accretion of material 
is the right answer. So far, our observations do not tell us anything about 
the collimation mechanism except that it must be very efficient: in the 
southern jet of OH\,231.8+4.2 the collimation factor is $\sim$~20; a similar 
number is found for the present post-AGB mass loss in M\,1--92.

\section{89\,Her}

This object is the prototype of the so called luminous high-latitude stars 
(LHL), which are though to represent the post-AGB phase of low-mass stars, 
although maybe it cannot be considered as a low-mass PPN, since it is not 
clear whether this type of objects will evolve into PNe. 89\,Her is a F2Ibe 
star (7\,000 K) with a luminosity of 3\,300~$L_{\sun}$ for an estimated 
distance of 600 pc (see Alcolea \& Bujarrabal 1995 and references therein). 
The star is a regular pulsator with a period of 60~d and, based on an 
additional 288 d period of its radial velocity, it is believed to be a close 
binary system (Waters et al$.$ 1993). The star was also known to present a 
strong IR 
excess due to a dusty circumstellar envelope, and a peculiar CO profile 
consisting of a very narrow component (FWHP $\sim$~3.5~km\,s$^{-1}$) plus 
weaker wings (FWZP $\sim$~20~km\,s$^{-1}$).

We observed this object with PdB in 1994--1995 using 4 antennas and
in the 3~mm band only. We mapped the $^{12}$CO $J$=1--0 line emission with
a spatial resolution of $\sim$~4$''$ (2$''$ when using uniform weighting). 
The results of these observations are that the molecular envelope consists of 
two components: a compact core, only partially resolved by the observations, 
and a detached shell. Both components show, in first approximation, spherical 
symmetry, in spite of that 89\,Her is a close binary system
(see Alcolea \& Bujarrabal 1995). This result does
neither support nor rejects binarity as responsible for the bipolarity in PNe,
but certainly supports the idea that maybe high-latitude (low-mass) objects 
are less likely to show strong deviations from spherical symmetry 
(Corradi \& Schwarz 1995)

\acknowledgments

This work has been partially supported by the Spanish DGES project PB96-0104.

\end{document}